\documentclass[prl,aps,amssymb,amsfonts,amsmath,showpacs,twocolumn]{revtex4}
\usepackage{graphicx}

\begin{document}

\title{On-command enhancement of single molecule fluorescence \\using a gold nanoparticle as an optical nano-antenna}

\author{Sergei~K\"{u}hn}
\author{Ulf~H\aa kanson}
\author{Lavinia~Rogobete}
\author{Vahid~Sandoghdar}\email{vahid.sandoghdar@ethz.ch}

\affiliation{Laboratory of Physical Chemistry, ETH Zurich, CH-8093
Zurich, Switzerland}

\begin{abstract}
We investigate the coupling of a single molecule to a single
spherical gold nanoparticle acting as a nano-antenna. Using
scanning probe technology, we position the particle in front of
the molecule with nanometer accuracy and measure a strong
enhancement of more than 20 times in the fluorescence intensity
simultaneous to a 20-fold shortening of the excited state
lifetime. Direct comparison with three-dimensional calculations
allow us to decipher the contributions of the excitation
enhancement, spontaneous emission modification, and quenching.
Furthermore, we provide direct evidence for the role of the
particle plasmon resonance in the modification of the molecular
emission.
\end{abstract}


\maketitle

\onecolumngrid
\begin{center}
submitted to Phys.Rev.Lett. 12/04/2005
\end{center}
\twocolumngrid

Metals are notorious for quenching the radiation of emitters
placed in their near field~\cite{chance:78,xie:94,Dulkeith:02}. On
the other hand, it is known that the
luminescence~\cite{Shimizu:02,Farahani:05} and Raman signals can
be enhanced on metallic
nanostructures~\cite{Nie:97,Kneipp:97,stoeckle:00,hayazawa:01}.
The desire to understand and exploit these phenomena has triggered
a large number of investigations over more than three
decades~\cite{Moskovits:85,Hartschuh:04,vanDuyne:04}. However,
quantitative measurements and comparisons with theoretical
predictions have been plagued by the lack of control on a large
number of parameters that determine the radiation properties of an
emitter close to a nanostructure. To address this challenging
issue, we have studied the interaction of a single oriented
molecule with a single spherical gold nanoparticle under
\emph{in-situ} position control (see Fig.~1a).

A gold nanoparticle (gnp) supports plasmon resonances associated
with the excitation of a collective oscillation of electrons. The
scattering properties and the plasmon spectra of small gnps are
well described in the quasistatic dipole approximation limit of
Mie theory if one takes into account radiation
damping~\cite{Kreibig,Kalkbrenner:04}. Thus, given a dipolar
radiation pattern and a well-defined resonance spectrum, a gnp
behaves as an elementary resonant dipole antenna. In what follow,
we investigate the strong influence of such a nano-antenna on the
excitation and emission of a single molecule (SM).

Let us consider an emitter with ground and excited states
$\left\vert g\right\rangle $ and $\left\vert e\right\rangle $
placed at the origin. The fluorescence signal detected from the
emitter is given by $S_f= \xi \sigma_{ee} \gamma_r$ where $\xi$ is
the overall detection efficiency of the setup, $\sigma_{ee}$ is
the population of the excited state, and $\gamma_r$ is its
radiative decay rate. To obtain $\sigma_{ee}$, one should consider
the effects of saturation and triplet bottle neck but in the
regime well below saturation, where all measurement in this work
were performed, $S_f \propto \xi \eta \left\vert \left\langle
e\left\vert \mathbf{E}.\mathbf{D}\right\vert g\right\rangle
\right\vert ^{2}$. Here $\eta=\gamma_r/(\gamma_r+\gamma_{nr})$
denotes the fluorescence quantum efficiency whereby $\gamma_{nr}$
is the nonradiative decay rate of $\left\vert e\right\rangle $.
The excitation electric field at the location of the molecule is
represented by $\mathbf{E}$, and $\mathbf{D}$ stands for molecular
dipole moment operator. Now if we introduce a nanostructure at
$\mathbf{r}=(x,y,z)$ with $r$ being much smaller than the
transition wavelength, the molecule experiences an inhomogeneous
$\mathbf{E}$ field so that the excitation rate becomes a very
sensitive function of $\mathbf{r}$ and the molecular orientation.
Moreover, the vicinity of the nanostructure alters both $\gamma_r$
and $\gamma_{nr}$ in a strongly distance and orientation dependent
manner~\cite{Ruppin:82,Das:02,Thomas:04}. To make the matter more
complicated, all these processes depend on the correspondence
between the possible plasmon resonances of the nanostructure and
the molecular excitation and emission wavelengths $\lambda_{exc}$
and $\lambda_f$, respectively. For a given nanostructure and a
molecular dipole moment of certain orientation we rewrite $S_f$
as:
\begin{equation}\label{fluo}
S_f \propto \xi I_0 d^2_{eg} \eta(\mathbf{r},\lambda_{f})
K(\mathbf{r},\lambda_{exc})
\end{equation} where $K$ represents
the excitation process and accounts for the enhancement of the
electric field intensity near the nanostructure as well as its
projection onto the direction of $\mathbf{D}$. The quantities
$I_0$ and $d_{eg}$ stand for the incident excitation intensity in
the absence of the nanostructure, and for the matrix element
associated with the $g$-$e$ transition.

\begin{figure}[b!]
\centering
\includegraphics[width=7.5 cm]{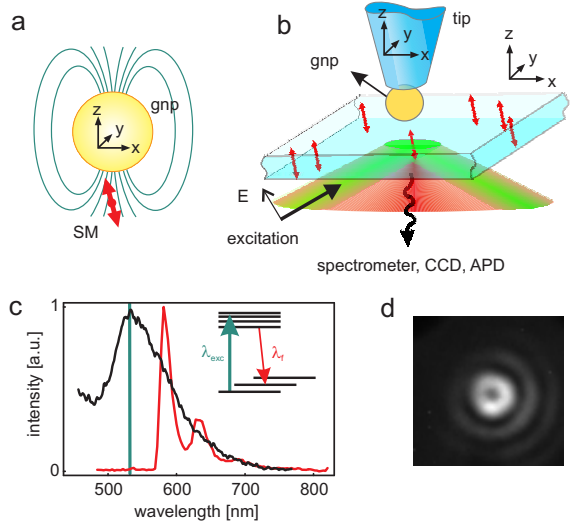}
\caption{(Color) a) The main idea of our work: a single gold
nanoparticle (gnp) is scanned across a single molecule (SM). The
green lines sketch the inhomogeneous excitation field. b) The
schematics of the experimental arrangement. c) The black curve
shows the measured plasmon spectrum of a gnp attached to the glass
fiber tip. The red curve displays the emission spectrum of
terrylene. The sharp short wavelength edge is due to the cut-off
filter. The green line marks the excitation wavelengths at 532~nm.
d) A CCD camera image of an SM.} \label{setup}
\end{figure}

The experimental setup consists of a scanning shear-force
stage~\cite{Kalkbrenner:01,Kalkbrenner:04,Buchler:05} mounted on
an inverted optical microscope (see Fig.~1b). Uncoated heat-pulled
fiber tips carrying single gnps (diameter 100~nm) at their
extremities were prepared according to the recipe reported in
Ref.~\cite{Kalkbrenner:01}. The plasmon spectrum of the gnp
attached to the tip was monitored by dark-field illumination with
white light from a xenon lamp through the microscope
objective~\cite{Kalkbrenner:04}. The black curve in Fig.~1c
displays the plasmon resonance of a typical gnp. The sample was a
20-30~nm thin para-terphenyl (pT) crystalline film doped with a
very low concentration ($10^{-9}$ molar) of terrylene molecules,
obtained from spin coating~\cite{pfab:04}. An important advantage
of this system is that the pT matrix is thin enough to allow
near-field studies. Furthermore, terrylene molecules are oriented
almost parallel to the $z$-axis, possess near unity quantum
efficiency and are remarkably
photostable~\cite{pfab:04,Buchler:05}. In addition, as indicated
in Fig.~1c, the excitation wavelength and fluorescence spectra of
terrylene nicely overlap with the gnp plasmon resonance, allowing
us to investigate the role of the latter in the modification of
both excitation and emission enhancement processes.

Terrylene molecules were excited by a pulsed laser at a wavelength
of $\lambda_{exc}=532$~nm with a pulse width of less than 30~ps. A
$p$-polarized laser beam was offset from the center of the
microscope objective (NA=1.4) so as to achieve total internal
reflection at the pT-air interface and to adapt the polarization
of the excitation light to the molecular dipole orientation in the
film. A sensitive CCD camera was used to identify individual
fluorescent molecules in a wide-field image of the
sample~\cite{pfab:04}. Figure~1d shows an example of an SM image.
The slight asymmetry in the doughnut-shaped pattern allows us to
determine the orientation of terrylene molecules in pT films to be
about $15\pm 5^{\circ}$ with respect to the substrate
normal~\cite{Buchler:05}. In what follows we take the position of
a given molecule to be at $x=y=0$ and $z=z_0$ whereby the
sample-air interface is set at $z=0$. To couple the gnp to single
terrylene molecules, the tip was approached to the pT sample and
distance stabilized at a separation of the order of 1~nm using
shear-force control. The gnp was next scanned laterally across
various molecules, allowing us to visualize directly a strong
increase of the molecular fluorescence on the camera. Then an SM
was selected and positioned at the center of the field of view,
and its fluorescence light was directed through a pinhole onto an
avalanche photodiode (APD) in a confocal detection arrangement.
The output of the APD was fed to a time card which was
synchronized with the arrival time of the laser pulses. By
recording the arrival time of each photon with respect to the
laser pulse, we determined the excited state decay time ($\tau$)
of the molecule. For terrylene $\tau$ is about 4~ns in bulk pT,
but it increases for molecules with perpendicular orientation
close to the pT-air interface~\cite{Buchler:05}. In our samples we
measured $\tau_0\simeq 20$~ns in the absence of the gnp.

\begin{figure}[b!]
\centering
\includegraphics[width=7.5 cm]{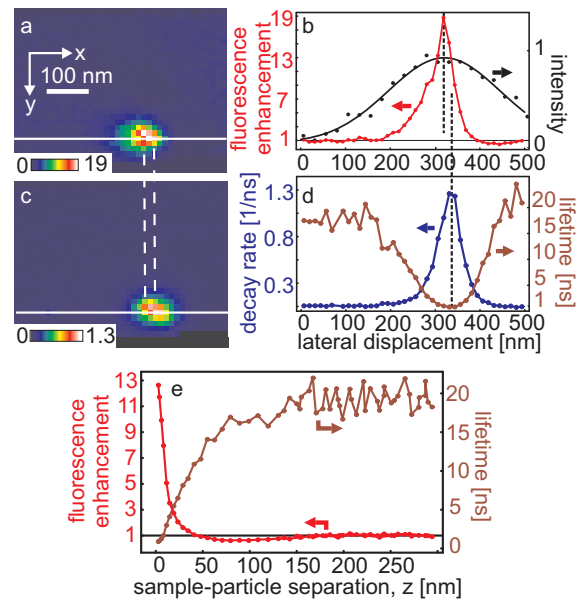}
\caption{(Color) a) Normalized fluorescence signal of an SM as a
function of the lateral position of the gnp placed nearly in
contact with the sample. b) The red curve shows a cross section
from a) after subtraction of a small background fluorescence. The
black curve displays, for comparison, a confocal fluorescence
signal as the molecule was scanned through a focused laser beam
(without the tip). c) Lateral position dependence of the
fluorescence decay rate recorded simultaneously as in image a). d)
The brown and blue curves show the fluorescence lifetime and the
corresponding decay rate for the same cross section shown in b).
e) The red curve represents $S_f/S^0_f$ as a function of $z$. The
brown curve shows $\tau$ measured simultaneously as $S_f$.}
\end{figure}

Figure~2a shows $S_f$ from an SM under constant illumination, as
the gnp was scanned laterally nearly in contact with the sample
surface. In other words, the blue background represents the
fluorescence signal $S^0_f$ of the molecule in the absence of the
gnp. In this measurement, the molecule was photobleached shortly
after the gnp passage~\cite{photostability}, allowing us to
determine the small residual fluorescence of the system. Figure~2b
displays $S_f/S^0_f$ for a cross section of Fig.~2a after
subtracting this residual fluorescence~\cite{goldfluo}. We find
that in this run $S_f$ is enhanced by up to 19 times when the gnp
is placed in the near field of the molecule.

Figure~2c displays the map of the excited state decay rate
$\gamma=1/\tau$ obtained from the measurements of $\tau$ performed
simultaneously as $S_f$ presented in Fig.~2a. The blue (brown)
curve in Fig.~2d shows $\gamma$ ($\tau$) from the same cut as in
Fig.~2a. Here we find that $\gamma$ increases by up to 22 times
when the gnp was scanned over the molecule. However, note that as
marked by the vertical dashed lines in Figs.~2a-d, the excellent
signal-to-noise ratio of the data reveals a small lateral shift of
about 20~nm between the maxima of $S_f$ and $\gamma$. Furthermore,
there is a slight asymmetry in the curves of Figs.~2b and 2d.
These features are due to the sensitive dependence of $S_f$ and
$\gamma$ on the relative orientation and position of the molecular
dipole with respect to the gnp.

\begin{figure}[b!]
\centering
\includegraphics[width=7.5 cm]{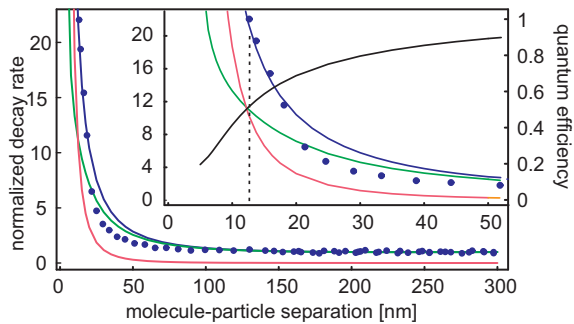}
\caption{(Color) The green, pink and blue curves represent the
theoretical values of radiative, nonradiative and total decay
rates respectively, as normalized to the unperturbed decay rate
$\gamma^0$. The symbols plot $\gamma=1/\tau$ from the data in
Fig.~2e for direct comparison. The inset shows a zoom of the
closer region. The black curve displays the resulting quantum
efficiency, the dashed line marks the closest molecule-gnp
separation.}
\end{figure}

In addition to scanning the particle in the \emph{xy} plane at
$z\simeq0$, we have also performed \emph{xyz }scans. This is
particularly useful for a quantitative analysis (discussed below)
because as opposed to scans along $x$ or $y$, the relative
orientation of the molecular dipole with respect to the gnp
remains constant during a $z$-scan. The red and brown curves in
Fig.~2e display $S_f/S^0_f$ and $\tau$ respectively, as a function
of $z$ (for a molecule different from the one in Figs.~2a-d). In
this run a 13-fold enhancement of $S_f$ is accompanied by a 22
times decrease (increase) of $\tau$ ($\gamma$) at the shortest
molecule-gnp separation. We point out, in passing, the slight dip
of the normalized fluorescence signal below one, which also occurs
in Fig.~2e. Our calculations hint that this is due to the
interference between the incident laser beam and its scattering
from the gnp.

The full width at half-maximum of the "image" of the molecule in
Figs.~2b and 2d is about 65~nm. For comparison, the black curve in
Fig.~2b shows a typical confocal fluorescence scan of a molecule
without the gnp. The near-field improvement of the resolution is
evident. The data in Fig.~2e illustrate that the enhancement
effects are even more confined in the $z$ direction, to only about
10~nm. In the context of optical microscopy, our results
demonstrate the realization of the most elementary "apertureless"
near-field optical microscope~\cite{zenhausern:94} where the probe
is a well-defined nanoscopic scattering center and the sample is a
single molecule. Reducing the size of the gnp would improve this
resolution further.

\begin{figure}[b!]
\centering
\includegraphics[width=7.5 cm]{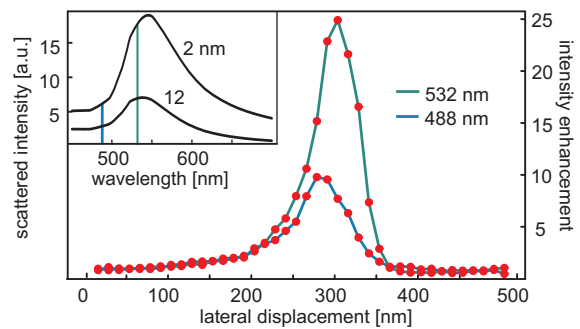}
\caption{(Color) Green and blue curves connecting red symbols show
the dependence of the fluorescence enhancement $S_f/S^0_f$ on the
lateral molecule-gnp displacement for the same single molecule
excited at $\lambda=532~nm$ and $488~nm$, respectively. The inset
shows the theoretical near-field intensity enhancement as a
function of wavelength for molecule-gnp separations of 2 and
12~nm.}
\end{figure}

To compare our data on the modification of the fluorescence decay
rate with theory, we have followed the formalism discussed in
Ref.~\cite{Das:02} to perform three-dimensional calculations of
$\gamma_r$ and $\gamma_{nr}$ for a molecule tilted by $15^{\circ}$
with respect to the radial axis of a gnp. We used the values of
the dielectric functions of gold at~$\lambda_{f}=580~nm$ provided
by Ref.~\cite{JohnsonChristy:72}. Figure~3 shows the change of
$\gamma_r$ (green), $\gamma_{nr}$ (pink) and the total decay rate
$\gamma=\gamma_r + \gamma_{nr}$ (blue), all normalized to the
unperturbed decay rate $\gamma^0$ and plotted as a function of the
molecule-gnp separation along the z-axis. To accommodate a direct
comparison with the experimental data, we have also included
$\gamma$ from Fig.~2e (blue symbols) and have fitted the blue
curve to the data points, leaving only the depth of the molecule
in the pT film ($z_0$) as a free parameter. The best fit yielded
$z_0=-12.5~nm$ for this molecule, which is well in the range of
our findings from independent measurements on similar
samples~\cite{Buchler:05}. The inset zooms into the short distance
region for a more quantitative scrutiny. The experimental and
theoretical values of $\gamma$ match very well except for a
deviation in a region of 20-50~nm. Preliminary two-dimensional
boundary integral calculations suggest that this discrepancy is
due to the effect of the pT-air interface~\cite{Rogobete:06}.

The black curve in the inset of Fig.~3 displays the calculated
$\eta$ which tends to zero at very small distances while
$\gamma_{nr}$ shoots up rapidly. Considering the very good
agreement between the theoretical and experimental values of
$\gamma$, we rely on the data in the inset to deduce $\eta\simeq
0.5$, $\gamma_r\simeq 11\gamma^0$ and $\gamma_{nr}\simeq
11\gamma^0$ for the closest molecule-gnp separation. Remembering
that $S_f$ has risen by about 13 times for the same measurement
(Fig.~2e), Eq.~(1) lets us conclude that $K\approx 25$ at this
separation. We emphasize that $\xi$ in Eq.~(1) does not undergo a
substantial change as the gnp is scanned over the molecule. First,
by performing two-dimensional calculations we have confirmed that
a small round nanoparticle does not change the emission pattern of
a radially oriented dipole in its near field. Furthermore, we have
calculated that our microscope objective collects more than $75\%$
of the fluorescence from a single molecule in a pT film supported
by glass. Therefore, any redirection of fluorescence by the gnp
would not result in a considerable increase of the detected
signal. To estimate the theoretically expected field intensity
enhancement $K$, we have used exact Mie theory to compute the near
field of a gnp illuminated by a plane wave~\cite{Kreibig}. The
inset in Fig.~4 shows examples of the near-field enhancement for
two gnp-molecule separations of $2$ and $12$~nm. The predicted
enhancement of about 5-20 is somewhat lower than the observed
value of $K$. Considering the strong distance dependence of the
excitation and emission processes close to the molecule, the
apparent surplus of enhancement could be caused by a small
inaccuracy in the assignment of $z_0$ or in the calculation of the
near-field enhancement. To eliminate these sources of error, we
require three-dimensional calculations that take into account the
effect of the pT-air interface both for the illumination and
fluorescence processes. These go beyond the scope of this work and
remain a subject of future investigations.

We have verified that replacing the gnp with an extended metallic
tip without a resonant character results in complete fluorescence
quenching instead of enhancement~\cite{Kuehn:06b}. In order to
provide a direct proof of the central role of the gnp plasmon
resonance in the enhancement process, we have examined the
fluorescence of the same single molecule under excitation at
wavelengths of 532~nm and 488~nm. As displayed in the inset of
Fig.~4, the near fields at these two wavelengths differ by a
factor of about 2.5. In Fig.~4 the red symbols connected by the
green and blue lines show the normalized intensity $S_f/S^0_f$ as
a function of the gnp position in the x-direction for the
excitation wavelengths 532 and 488~nm, respectively. We find that
$S_f$ is indeed enhanced by about 2.5 times more under green
excitation, illustrating the importance of the antenna resonance
in its interaction with the molecule. Our measurements of the
molecular emission spectrum under a gnp (not shown here) also
clearly reveal the influence of the plasmon spectrum on
$\gamma_r$.

In conclusion, we have demonstrated a quantitative and controlled
enhancement of single molecule fluorescence due to its near-field
coupling with a gold nanoparticle. Moreover, we have presented a
clear evidence of the role of local plasmon resonances in the
excitation process. The choice of a well-characterized system and
the quality of our measurements made it possible to compare the
data directly with three-dimensional calculations for individual
molecules, without the need for statistical averaging. We
emphasize that the experimental procedure and observations
presented here could be reproduced routinely. Our approach based
on the on-command manipulation of a nanoparticle acting as a
nano-antenna is also promising for various applications such as
detection of weakly fluorescing systems~\cite{Barnes:00}.

We thank R. Carminati, I. Gerhardt, and A. Imamoglu for fruitful
discussions. This work was supported by the Swiss National
Foundation and the Swiss Ministry of Education and Science (EU
IP-Molecular Imaging).


\begin{thebibliography}{26}
\bibitem[{pho()}]{photostability}
\bibinfo{note}{Although terrylene molecules are extremely photostable in this
  system, we have observed that the presence of the tip very close to the
  molecule results in a limited photostability. This phenomenon is currently
  under investigation and will be addressed in a future report.}
\bibitem[{gol()}]{goldfluo}
\bibinfo{note}{When the gnp enters the evanescent field of the excitation light
  at the air-pT interface, it fluoresces by a very small amount. However, since
  this signal is weak and constant, we can account for it without any
  complications.}
\end{thebibliography}

\end{document}